\documentclass[preprint,showpacs,preprintnumbers,amsmath,amssymb]{revtex4}

\usepackage{graphicx}
\usepackage{dcolumn}
\usepackage{bm}
\usepackage{here}

\begin{document}

\preprint{Revtex4}

\title{Electronic Localization Properties of a Double Strand of DNA:
A Simple Model with Long-Range Correlated Hopping Disorder
}

\author{Hiroaki Yamada}
 \email{hyamada@cc.niigata-u.ac.jp}
\affiliation{%
Aoyama 5-7-14-205, Niigata 950-2002, Japan 
\footnote{Formerly address: Faculty of Engineering,  Niigata University,
Ikarashi 2-Nochou 8050, Niigata 950-2181, Japan} 
}%
\date{\today}

\begin{abstract}
Localization property in the disordered few-chain DNA systems with a long-range correlation
is numerically investigated.
We apply the chain system with the correlated disorder 
in the interchain and/or intrachain hoppings
to the simple model of a double strand of DNA.
Numerical results for the density of states and 
the Lyapunov exponent of 
the wave function in the two- or three-chain models are given.
It is found that the correlation effect enhances the localization 
length (the inverse least nonnegative Lyapunov exponent) around the band center.
\end{abstract}

\maketitle

\section{Introduction}
Localization phenomena in the one-dimensional disordered system
have been extensively studied \cite{lifshiz88}.  
It is well known that almost all the
eigenstates are exponentially localized and 
the system has a pure point energy spectrum 
under the presence of any disorder \cite{ishii73}.
 Recently, the effect of the long-range correlated disorder in the potential
field on the localization 
has been reported by some groups 
\cite{yamada91,moura98,carpena02,yamada03,yamada03a}.
 The appearance of the Anderson-like metal-insulator transition is suggested 
in the strongly correlated non-stationary regime of the sequence 
\cite{moura98}.
 As one of the realistic situation, it has been found that the base (nucleotide)
sequence of the various genes 
has a strong correlation 
characterized by the power spectrum 
$S(f)\sim f^{-\alpha}$ ($0.2<\alpha <0.8$)
\cite{voss92,bishop97,holste01,holste03,carpena02,isohata03,grosse02}. 

Moreover, the recent development of the nanoscale fabrication let us expect
the utilization of the DNA wire as a molecular device \cite{lewis03,porath04} 
and the realization of DNA computing \cite{kari97,paun98}.
Actually, the development enables us
to measure the direct DNA transport phenomena \cite{porath00,tran00,zhong03,porath04}.
   Recently, Porth {\it et al.} measured the  
nonequilibrium current-voltage ($I-V$) characteristics in the
poly(G)-poly(C) DNA molecule attached to platinum lead at room temperature 
\cite{porath00}. 
 Cuniberti {\it et al.} explained the semiconducting behavior by considering
the base-pair stack coupled to the sugar-phosphate(SP) backbone pair
\cite{cuniberti02}.
 Iguchi also derived the semiconductivity and the band gap by using 
the ladder chain model of the double strand of DNA \cite{iguchi97,iguchi01,zhong03}.
 In the both models, apparently, the existence of the SP backbone chain  
play an important role in the band structure due to 
the gap opening
 by the hybridization of the energy levels.

 On the other hand, 
recently Tran et al measured the conductivity along the lambda phage DNA ($\lambda-$DNA)
 double helix at microwave frequencies using the lyophilized DNA in and also 
without a buffer \cite{tran00}. 
The conductivity is strongly temperature dependent around room temperature with 
a crossover to a weakly temperature dependent conductivity at low temperatures.
 Yu and Song showed that the observed temperature dependent
conductivity in the DNA can be consistently
modeled, without invoking the additional ionic conduction
mechanisms, by considering that electrons may use
the variable range hopping for conduction and that electron
localization is enhanced by strong thermal structural
fluctuations in DNA \cite{yu01}.
  Then the DNA double helix is viewed as a one-dimensional Anderson system.

As observed in the power spectrum, the mutual information analysis 
and the Zipf analysis of the DNA base
sequence such as the human chromosome 22, the long-range structural correlation exists in the 
total sequence as well as the short-range periodicity \cite{holste01,holste03,isohata03,grosse02}.   
 The transport property though DNA are still controversial mainly
due to the complexity of the experimental environment
and the molecule itself.
 Although the theoretical explanations for the phenomena have been tried 
by some standard pictures used in the solid state physics such as 
polarons, solitons, electrons or holes 
\cite{iguchi03,iguchi03a,hennig02,hennig03,hennig03a,ladik99,shen02,bruinsma00,kats02,starikov02,damle02,troisi02,rodriguez03,lewis03},
the situation is still far from unifying the theoretical scheme.

In the present paper we investigate the correlation effect on 
the localization property of the one-electronic states 
in the disordered,  two-chain (ladder) \cite{yamada04} and three-chain models  
with a long-range structural correlation
as a simple model for the electronic property in the DNA.
  The tight-binding model for 
the few-chain systems have the off-diagonal randomness as the interchain 
hopping integral and/or the intrachain ones.
We present some numerical results for 
the density of states (DOS) and 
the Lyapunov exponents of the wave function.
 In particular, it is found that the correlation of the sequence enhances the 
localization length defined by using the least nonnegative Lyapunov exponent.
Regardless of the parameter tuning for the numerical calculation,
we would like to mainly focus on (1)suggesting the model and (2)giving
the preliminary numerical results for the correlation effect on
the localization in the model.

  Outline of the present paper is as follows.
In the next section we introduce the models for DNA with a long-range correlation. 
In the Sect.3 we give the results for the density of states. 
The numerical results for the Lyapunov exponent and the localization length 
in the systems are given in Sect.4. 
The last section devotes for summary and discussion.

\section{Model}
We simplify and model the double strand of DNA by some assumptions.
DNA double helix structure is constructed by the coupled two single strand of 
DNA.
 First, we ignore the twist of DNA as well as the complicated topology. 
(See Fig.1.)
In addition to the simplification, we consider only the $\pi-$electrons with the
highest occupied molecular orbit (HOMO) states in the backbones and 
the base-pairs of the system. We also ignore the interaction 
between the electrons and restrict ourselves to the zero-temperature property.

\begin{figure}[h]
\includegraphics[scale=0.5]{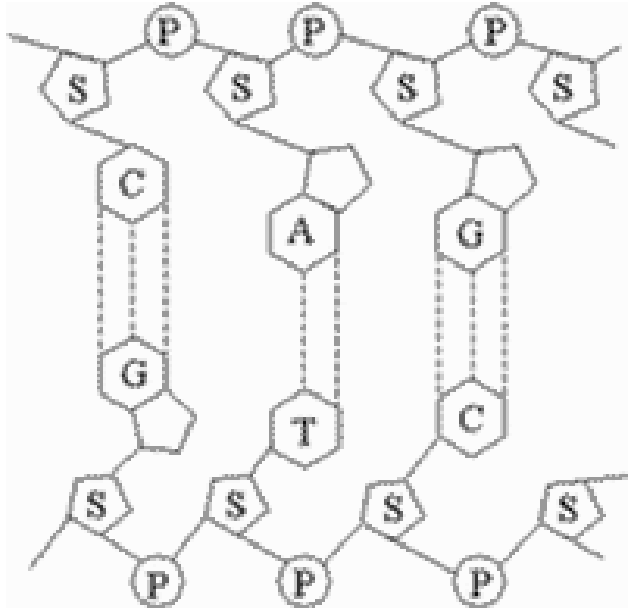}
  \caption{The schematic structure of a double strand of DNA.
 }
\end{figure}

\subsection{Two-chain model}
Following the basic assumptions, 
consider the one-electron system described by  
the tightly binding Hamiltonian $\hat{H}$ consisting of the 
two chains.
This model was first given by Iguchi \cite{iguchi97,iguchi01}
as a model for considering the electronic properties of
a double strand of DNA.
\begin{eqnarray}  
\hat{H} &=& H_A + H_B + H_{AB}, \\
H_k &=& \sum_{n} \{ k_{n,n}|k:n \rangle \langle k:n|  
 - (k_{n+1,n}|k:n+1 \rangle \langle k:n| + h.c. ) \}, \\
H_{AB} &=& -\sum_{n} V_n(|A:n \rangle \langle B:n| + 
 |B:n \rangle \langle A:n|),
\end{eqnarray}  
\noindent
where the $\{|A:n>,|B:n> \}$ denotes an orthonormalized set and 
$A_{n+1,n}$ ($B_{n+1,n}$) means the hopping integral 
between the $n$th and $(n+1)$th sites and $A_{n,n}$ ($B_{n,n}$) 
the on-site energy at site $n$ in chain $A$
($B$), and $V_{n}$ is the hopping integral 
from chain $A (B)$ to chain $B (A)$ at site $n$,
respectively.  
   The Schr\"{o}dinger equation
$\hat{H}|\Psi\rangle = E|\Psi\rangle$ becomes,
\begin{eqnarray*}  
A_{n+1,n} \phi^{A}_{n+1}+ A_{n,n-1} \phi^{A}_{n-1} + A_{n,n} \phi^{A}_{n} 
+ V_{n} \phi^{B}_{n}  = E\phi^{A}_{n}, \\ 
B_{n+1,n} \phi^{B}_{n+1}+ B_{n,n-1} \phi^{B}_{n-1} + B_{n,n} \phi^{B}_{n} 
+ V_{n} \phi^{A}_{n}  = E\phi^{B}_{n},  
\end{eqnarray*}  
\noindent
where $\phi^{A}_{n} \equiv \langle A:n|\Psi\rangle$ and 
$\phi^{B}_{n} \equiv \langle B:n|\Psi\rangle$. 
Furthermore it can be rewritten in the matrix form,
\begin{eqnarray}  
 \left(
\begin{array}{c}
\phi^{A}_{n+1} \\
\phi^{B}_{n+1} \\ 
\phi^{A}_{n} \\ 
\phi^{B}_{n} 
\end{array}
\right)
=
\left(
\begin{array}{cccc}
\frac{E-A_{nn}}{A_{n+1n}} & -\frac{V_n}{B_{n+1n}} & -\frac{A_{n-1n}}{A_{n+1n}} & 0 \\
-\frac{V_n}{B_{n+1n}}  & \frac{E-B_{nn}}{B_{n+1n}} & 0& -\frac{B_{n-1n}}{B_{n+1n}}  \\
 1 & 0 & 0 & 0 \\
0 & 1 & 0 & 0 
\end{array}
\right)
\left(
\begin{array}{c}
\phi^{A}_{n} \\
\phi^{B}_{n} \\ 
\phi^{A}_{n-1} \\ 
\phi^{B}_{n-1} 
\end{array}
\right) 
\equiv T(n)
\left(
\begin{array}{c}
\phi^{A}_{n} \\
\phi^{B}_{n} \\ 
\phi^{A}_{n-1} \\ 
\phi^{B}_{n-1} 
\end{array}
\right).
\end{eqnarray}  
\noindent
  Generally speaking, we would like to investigate the asymptotic 
behavior ($n \to \infty$) of the products of the matrices 
$M_n=\Pi_k^n T(k)$. 
According to the parameter sets given by Iguchi \cite{iguchi01},
we set $A_{n+1,n} = B_{n+1,n}=a(=b)$ at odd (even) site $n$, 
respectively and $V_n=0$ at even sites (phosphate sites)
 for simplicity. 
The chain $A$ and $B$ are constructed by the repetition 
of the sugar-phosphate sites, and the inter-chain hopping $V_n$
at the sugar sites come from the nucleotide base-pairs, i.e., $A-T$ or $G-C$ pairs.   
(See Fig.2(a).)

\begin{figure}[h]
\includegraphics[scale=1.0]{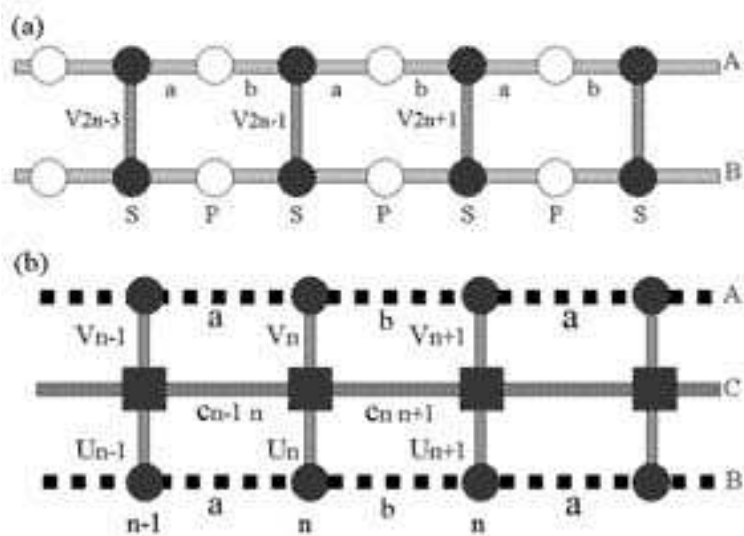}
  \caption{Models of the double starnd of DNA.
(a) the two-chain model, (b) the three-chain model.}
\end{figure}

\subsection{Three-chain model}
When we pay attention to the overlap integral between the nucleotide base-pairs, 
the two-chain model can be easily extended to the three-chain one described by following
Hamiltonian.
(See Fig.2(b).)
\begin{eqnarray}  
\hat{H} &=& H_A + H_B + H_C + H_{AC} + H_{BC}, \\
H_{AC} + H_{BC} &=& -\sum_{n} (V_n|A:n \rangle \langle C:n| + U_n|B:n \rangle \langle C:n| + h.c.),
\end{eqnarray}  
\noindent
where the $\{|A:n>,|C:n>,|B:n> \}$ denotes an orthonormalized set and 
$C_{n+1,n}$  means the hopping integral 
between the $n$th and $(n+1)$th sites and $C_{n,n}$  
the on-site energy at site $n$ in chain $C$, and $V_{n}$ 
and $U_n$ are the hopping integral between the chains.
  The Schr\"{o}dinger equation
$\hat{H}|\Psi\rangle = E|\Psi\rangle$ becomes,
\begin{eqnarray*}  
A_{n+1,n} \phi^{A}_{n+1}+ A_{n,n-1} \phi^{A}_{n-1} + A_{n,n} \phi^{A}_{n} 
+ V_{n} \phi^{C}_{n}  = E\phi^{A}_{n}, \\ 
C_{n+1,n} \phi^{C}_{n+1}+ C_{n,n-1} \phi^{C}_{n-1} + C_{n,n} \phi^{C}_{n} 
+ V_{n} \phi^{A}_{n} + U_{n} \phi^{B}_{n}   = E\phi^{C}_{n}, \\ 
B_{n+1,n} \phi^{B}_{n+1}+ B_{n,n-1} \phi^{B}_{n-1} + B_{n,n} \phi^{B}_{n} 
+ U_{n} \phi^{C}_{n}  = E\phi^{B}_{n},  
\end{eqnarray*}  
\noindent
where $\phi^{C}_{n} \equiv \langle C:n|\Psi\rangle$. 
It can be rewritten in the matrix form,
\begin{eqnarray}  
 \left(
\begin{array}{c}
\Phi_{n+1} \\
\Phi_{n} 
\end{array}
\right)
= T(n)
 \left(
\begin{array}{c}
\Phi_{n} \\
\Phi_{n-1} 
\end{array}
\right),
T(n)=
\left(
\begin{array}{cc}
H_d & H_o \\
I & O  
\end{array}
\right).
\end{eqnarray}  
\noindent
where $\Phi_{n}=(\phi^{A}_{n},\phi^{C}_{n},\phi^{B}_{n})^t$ and 
$I$ and $O$ denote the three-dimensional unit and zero matrices, respectively.
The matrices $H_d$ and $H_o$ are given as follows:
\begin{eqnarray}  
H_d=
 \left(
\begin{array}{ccc}
\frac{E-A_{nn}}{A_{n+1n}} & -\frac{V_n}{A_{n+1n}} & 0 \\
-\frac{V_n}{C_{n+1n}} &  \frac{E-C_{nn}}{C_{n+1n}} & -\frac{U_n}{C_{n+1n}} \\
0 & -\frac{U_n}{B_{n+1n}} & \frac{E-B_{nn}}{B_{n+1n}} 
\end{array}
\right), 
H_o=
 \left(
\begin{array}{ccc}
 -\frac{A_{n-1n}}{A_{n+1n}} & 0 &0 \\
 0 & -\frac{C_{n-1n}}{C_{n+1n}} & 0 \\
 0 & 0 & -\frac{B_{n-1n}}{B_{n+1n}} 
\end{array}
\right). 
\end{eqnarray}  
\noindent
In this paper we set $A_{n+1,n} = B_{n+1,n}=a(=b)$ at odd(even) site $n$,  
and $V_n=U_n$. 
(See Fig.2(b).) 
In addition, in the numerical calculation 
we set the on-site energy as $A_{nn}=B_{nn}=C_{nn}=0$ for simplicity.
 The sequence $\{ C_{nn+1} \}$ can be also generated by corresponding to the 
base-pairs sequence.
 The localization properties in the simple few-chain models with the on-site
disorder have been extensively investigated \cite{heinrichs02}.

\subsection{The correlated sequence}
The correlated binary sequence $\{V_n\}$ and/or $\{C_{nn+1} \}$ 
of the hopping integrals can be generated by 
 the modified Bernoulli map 
\cite{aizawa84,yamada91,yamada01,lahiri99}.

\begin{eqnarray}
 X_{n+1}= 
& \biggl( 
   \begin{array}{cc}
   X_{n} + 2^{B-1}X_{n}^{B}  & (X_{n} \in I_0)  \\
   X_{n} - 2^{B-1}(1-X_{n})^{B} & (X_{n} \in I_1),    
\end{array}
\label{eq:map}
\end{eqnarray}
\noindent
where $I_0=[0,1/2),I_1=[1/2,1)$.  
$B$ is a bifurcation parameter which controls the correlation
of the sequence. 

In the ladder model we use the symbolized sequence $\{ V_{n} \}$ 
 by the following rule as the interchain hopping integral at odd sites $n$:

\begin{eqnarray}
V_{n} =  
\left\{ \begin{array}{ll}
W_{AT} = W_{TA} & (X_{n} \in I_0) \\
W_{GC} = W_{CG} & (X_{n} \in I_1). \\
\end{array} \right.
\end{eqnarray} 
\noindent
In the numerical calculation, we use $W_{GC}=W_{AT}/2$ for simplicity.
Then the artificial binary sequence can be roughly regarded as the base-pair
sequence as observed in the $\lambda-$DNA or the human chromosome 22. 
The correlation function $C(n)(\equiv <V_{n_0}V_{n_0 + n}>)$ ($n_0=1$, $n$ is even ) 
decays by the inverse power-law depending on the value $B$ as 
$C(n)\sim n^{-\frac{2-B}{B-1}}$ for large $n$ ($3/2<B<2$). 
The power spectrum becomes 
$S(f)\sim f^{-\frac{2B-3}{B-1}}$ for small $f$. 
 We focus on the Gaussian ($1< B< 3/2$) and the non-Gaussian stationary region 
$2/3\leq B<2$, corresponding to the DNA base-pairs sequence ($0.2< \alpha <0.8$).

  In the three-chain model we assume that the phosphate sites are renormalized 
in the hopping integral between the sugar sites for simplicity. 
 The interchain hopping integrals $V_n=U_n$ can be generated by the same way
to the two-chain model for every site $n$.
 Furthermore, we use the successive sequence $\{X_n,X_{n+1} \}$ 
when we make the correlated binary sequence 
$\{ C_{nn+1} \}$ as the hopping integral of 
the middle (nucleotide) chain as follows:

\begin{eqnarray}
C_{nn+1} =  
\left\{ \begin{array}{ll}
W_{AT-AT} & (X_{n} \in I_0, X_{n+1} \in I_0) \\
W_{AT-GC} & (X_{n} \in I_0, X_{n+1} \in I_1) \\
W_{GC-AT} & (X_{n} \in I_1, X_{n+1} \in I_0) \\
W_{GC-GC} & (X_{n} \in I_1, X_{n+1} \in I_1). \\
\end{array} \right.
\end{eqnarray} 
\noindent
In the numerical calculation we assume the following rules for simplicity.
\begin{eqnarray}
\left\{ \begin{array}{l}
W_{GC-GC} = W_{AT-AT}/2 \\
W_{AT-GC} = (W_{AT-AT} + W_{GC-GC})/2 \\
W_{GC-AT} =  W_{AT-GC}. \\
\end{array} \right.
\end{eqnarray} 
\noindent
As a result, the parameters are $W_{AT}$ and $W_{AT-AT}$.

\section{Density of States and Lyapunov Exponents} 
 In this section, we show the DOS and the Lyapunov exponents 
for the periodic and disordered 
two-chain model in order to give some notations and to 
confirm the numerical reliability. 

Let us consider the periodic case to confirm the gap opening
mechanism due to hybridization by the interchain coupling.
Figure 3 shows the DOS for various periodic cases.
As seen in the Fig.3(b) the energy spectrum consists of four bands when
$W_{AT}=W_{GC}$ and $A_{nn}=B_{nn}=0$.
The band gap $E_g(v)$ at the center is given by 
$E_g(v)=\surd(E_g^2+v^2)-v$, where $E_g\equiv 2(a-b)$ means the band gap
for the single strand of DNA. The other band gap 
$\Delta_g(v)=v+\frac{1}{2}[\surd (E_g^2+v^2)+\surd(4(a+b)^2+v^2)]$
appears in between the lowest and the lower middle bands where $v>2ab/\surd(a^2+b^2)$.
Iguchi showed the semiconductivity of the double strand of DNA based on the 
band structure \cite{iguchi01,iguchi03a}.
  The energy spectrum in the case $a=b$ is given in Fig.3(a).
The band gap at the center disappears.
  Furthermore, the energy band for case $W_{AT}\neq W_{GC}$($W_{GC}=W_{AT}/2$)
is shown in Fig.3(c). 
Each band splits and the total spectrum consists of eight bands.


Figure 4 shows the DOS as a function of the energy for the binary disordered system.
The sequence of the interchain hopping takes an alternative value 
$W_{GC}$ or $W_{AT}$. It is found that some gaps observed in the periodic 
cases close due to the disorder corresponding to the base-pairs sequence.

It is noted that 
in the bipartite lattice with an even number of sites the energy spectrum is 
striking symmetric around $E=0$ 
\cite{milke97,kozlov98,adame00,biswas00,brouwer98,brouwer02}. 
In this case, related to the chiral universality class,
 the $E=0$ states are non-localized in any dimension.
 However, the states with the Dyson 
singularity \cite{dyson53} at the center 
disappear when the randomness is introduced in the on-site
energy at the backbone chains.
  We comment the point in the next section again.

\begin{figure}[h]
\includegraphics[scale=0.7]{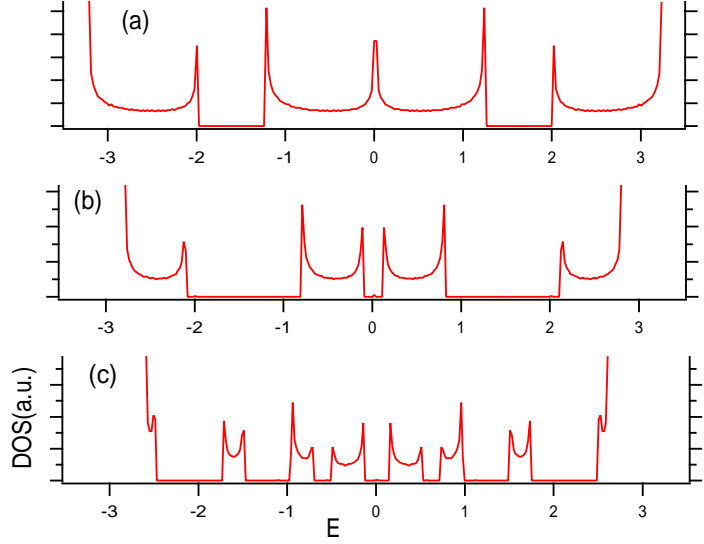}
  \caption{DOS as a function of energy for the periodic cases 
in the two-chain model. (a)$W_{AT}=W_{GC}=2.0, a=1.0, b=1.0$. 
 (b)$W_{AT}=W_{GC}=2.0, a=1.0, b=0.5$. 
 (c)$W_{AT}=2.0, W_{GC}=1.0, a=1.0, b=1.0$. 
The on-site energy is set at $A_{nn}=B_{nn}=0$.}
\end{figure}

\begin{figure}[h]
\includegraphics[scale=0.7]{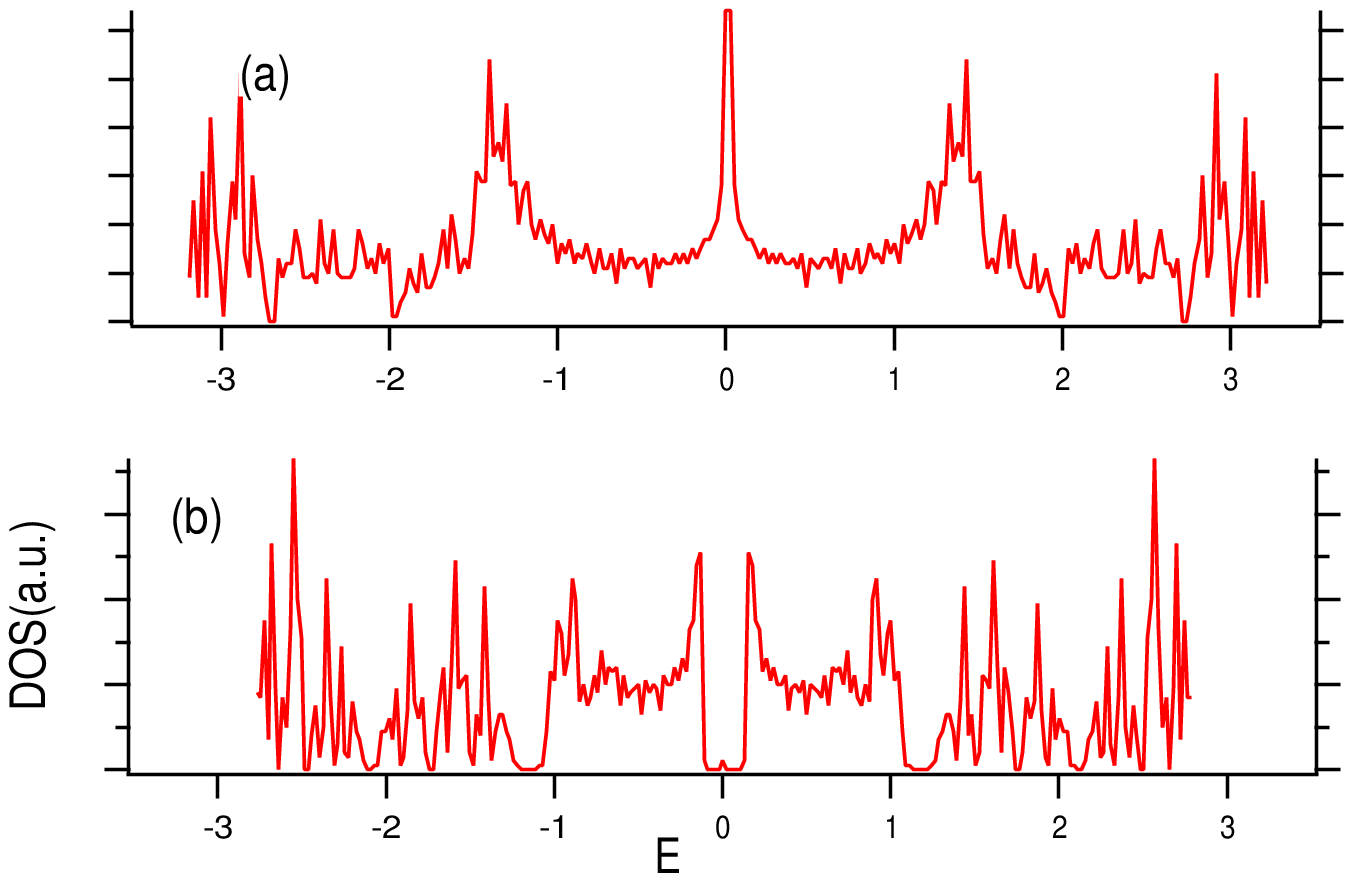}
  \caption{DOS as a function of energy for the binary disordered cases 
in the two-chain model.
 (a)$W_{AT}=2.0, W_{GC}=1.0, a=1.0, b=1.0$.
 (b)$W_{AT}=2.0, W_{GC}=1.0, a=1.0, b=0.5$. 
The on-site energy is set at $A_{nn}=B_{nn}=0$.}
\end{figure}

We give a preliminary numerical result of 
the energy dependence of the Lyapunov exponents.
The definition is given by,
\begin{eqnarray}  
\gamma_i = \lim_{n \to \infty} \frac{1}{2n} \log \sigma_i(M_n^\dagger M_n),    
\end{eqnarray}  
\noindent
where $\sigma_i(...) $ denotes the $i$th eigenvalue \cite{yamada01}. 
 As the transfer matrix $T(n)$ is symplectic, the eigenvalues of the
$M_n^\dagger M_n$ have the reciprocal symmetry around the unity as
$e^{\gamma_1},...,e^{\gamma_d},e^{-\gamma_d},...,e^{-\gamma_1}$, 
where $\gamma_1 \geq \gamma_2\geq \dots \gamma_d \geq  0$.
 The $d$ denotes the number of channels, i.e. $d=2$ in the two-chain model
and $d=3$ in the three-chain model.

\begin{figure}[h]
\includegraphics[scale=0.5]{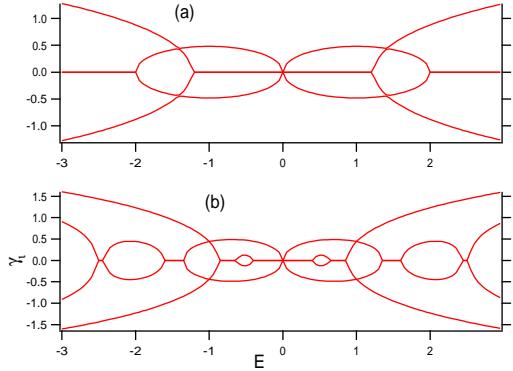}
  \caption{Lyapunov exponents as a function of energy for the periodic cases 
in the two-chain model. (a)$W_{AT}=W_{GC}=2.0, a=1.0, b=1.0$. 
 (b)$W_{AT}=2.0, W_{GC}=1.0, a=1.0, b=1.0$. 
The on-site energy is set at $A_{nn}=B_{nn}=0$.}
\end{figure}

Figure 5 shows all of the Lyapunov exponents $\gamma_i(i=1,2,3,4)$
as a function of the energy to confirm the reciprocal symmetry.
The energy regions where two exponents are positive 
($\gamma_1>\gamma_2>0 $) correspond to the energy gap in Fig.3 except for 
the vicinity of the singular point $E=0$.
  The zero Lyapunov exponent corresponds to the brunch cut in the 
analyticity of the energy spectrum.

The Lyapunov exponents are related to the DOS $\rho(E)$ as an analogue of the Thouless
relation as \cite{souillard86,molinari02},
\begin{eqnarray}
 \sum_i^d \gamma_i(E) \sim \int \ln |E-E^{'}| \rho(E^{'}) dE^{'}.   
\end{eqnarray}
\noindent
Note that the Thouless relation ($d=1$) is true only for the positive Lyapunov exponents. 
  One can see that 
the singularity of the largest Lyapunov exponent is strongly related to the 
singularity of the DOS.

Furthermore, it is found that 
for the thermodynamic limit 
the largest channel-dependent localization length $\xi_d=1/\gamma_d$ 
determines the exponential decay of the Landauer conductance $g$ 
which is in units of $e^2/h$ at 
zero temperature and serves as the localization
length of the total system of the coupled chains \cite{imry99,imry02}.
\begin{eqnarray}
 g(m) = 2 \sum_{i=1}^{d} \frac{1}{\cosh \frac{2m}{\xi_i(m)} -1} \sim \exp (-\frac{2m}{\xi_d(m)} ).
\end{eqnarray}
\noindent
Recently, electron transport for the molecular wire between two metalic electrorodes
has been also investigated by several techniques 
\cite{tikhonov02a,tikhonov02b,lehmann02,walczak03}.

\section{Correlation Effects}
 In this section, we consider the correlation effect on the localization
property of the disordered case by using the modified Bernoulli model.
 Figure 6 shows some eigenstates for the correlated case,
$B=1.8$. 
 Apparently the eigenstates are exponentially localized.
   In particular, the energy dependence of 
the Lyapunov exponents and the localization length are investigated 
in order to characterize the localization.
We used a large sample with the system size $N=10^6$ for the numerical
calculation.

\begin{figure}[h]
\includegraphics[scale=0.7]{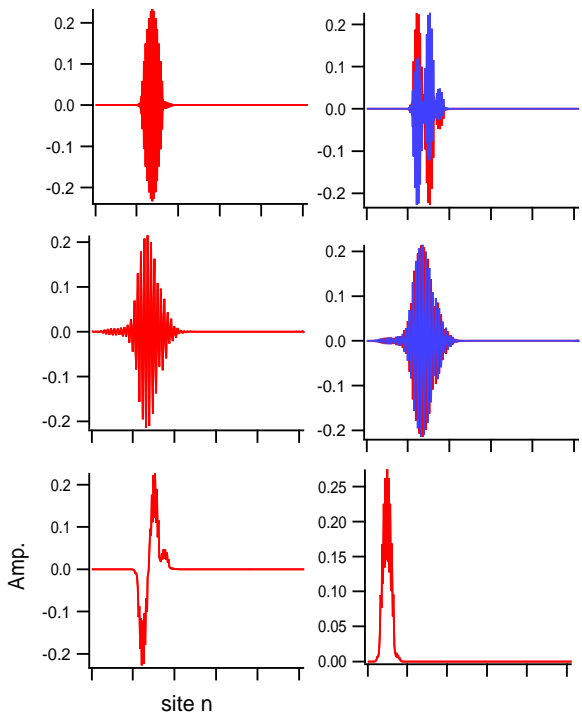}
  \caption{Some eigenfunctions for the binary correlated disordered cases $B=1.8$
in the two-chain model ($N=256$).
 The amplitude of the wave function for A and B chains are over-written.
$W_{AT}=2.0, W_{GC}=1.0, a=1.0, b=0.5$. 
The on-site energy is set at $A_{nn}=B_{nn}=0$.}
\end{figure}

\subsection{Two-chain model}
 First, we show the expectable results by a simple analytical
calculation.
We can write the Schr\H{o}dinger equation in the following matrix form
when the hopping integral and the on-site energy of the backbone chains 
are the same values ($a=b,A_{nn}=B_{nn}$),
\begin{eqnarray*}  
\left(
\begin{array}{c}
\phi^{A}_{n+1}+  \phi^{A}_{n-1} \\
\phi^{B}_{n+1}+  \phi^{B}_{n-1} 
\end{array}
\right)
=
\frac{1}{a}\left(
\begin{array}{cc}
E-A_{nn} & -V_n \\
-V_n & E-B_{nn}
\end{array}
\right)
\left(
\begin{array}{c}
\phi^{A}_{n} \\
\phi^{B}_{n} 
\end{array}
\right).
\end{eqnarray*}  
\noindent
It is easily seen that 
the equations can be decoupled to 
the one-dimensional 
Anderson model with diagonal randomness
by the unitary transform $U$ as follows;
\begin{eqnarray}  
 \left(
\begin{array}{c}
\psi^{A}_{n} \\
\psi^{B}_{n} 
\end{array}
\right)
=U
\left(
\begin{array}{c}
\phi^{A}_{n} \\
\phi^{B}_{n} 
\end{array}
\right)
,
U= \frac{1}{\sqrt 2}
\left(
\begin{array}{cc}
1 & 1 \\
1 & -1
\end{array}
\right).
\end{eqnarray}  
\noindent

\begin{eqnarray}  
\left(
\begin{array}{c}
 \psi^{A}_{n+1}+  \psi^{A}_{n-1} \\
 \psi^{B}_{n+1}+  \psi^{B}_{n-1} 
\end{array}
\right)
=
\frac{1}{a}\left(
\begin{array}{cc}
E-V_{n}-\mu_{n}  & \nu_{n} \\
\nu_{n} & E+V_{n}-\mu_{n} 
\end{array}
\right)
\left(
\begin{array}{c}
\psi^{A}_{n} \\
\psi^{B}_{n} 
\end{array}
\right),
\end{eqnarray}  
\noindent
where  
\begin{eqnarray}  
\mu_n=\frac{1}{2}(A_{nn}+B_{nn}), 
\nu_n=\frac{1}{2}(A_{nn}-B_{nn}).   
\end{eqnarray}  
\noindent

\begin{figure}[h]
\includegraphics[scale=0.5]{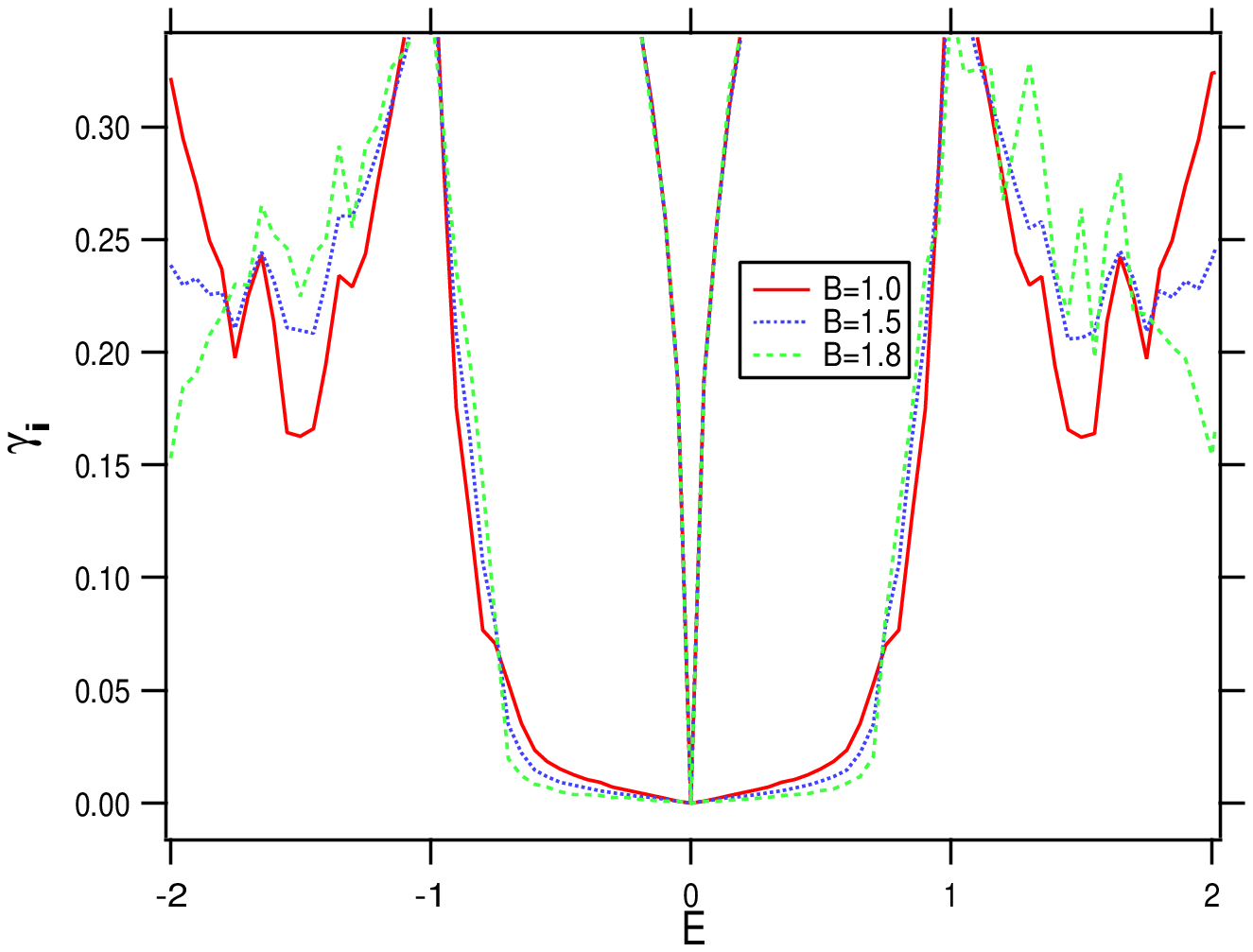}
  \caption{Lyapunov exponents $\gamma_i (i=1,2)$
  as a function of energy in the correlated two-chain model.
The parameters $W_{AT}=2.0, a=1.0, b=1.0$.
The on-site energy is set at $A_{nn}=B_{nn}=0$.
The number of the matrix product is $N=10^6$.}
\end{figure}

\begin{figure}[h]
\includegraphics[scale=0.7]{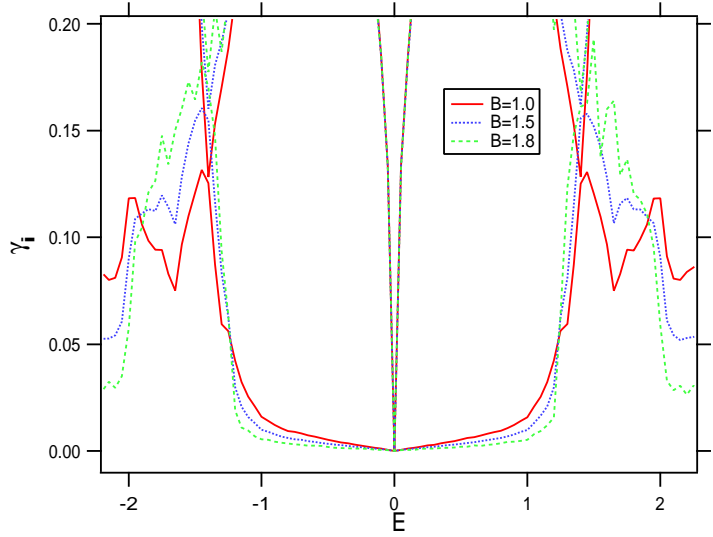}
  \caption{Lyapunov exponents $\gamma_i (i=1,2)$
  as a function of energy in the correlated two-chain model.
The parameters $W_{AT}=2.0, a=1.0, b=0.5$.
The on-site energy is set at $A_{nn}=B_{nn}=0$.
The number of the matrix product is $N=10^6$.}
\end{figure}

In our case, $\mu_n=A_{nn}, \nu_n=0$.
Accordingly, the behavior of the Lyapunov exponents are essentially the same to that 
in the one-dimensional system with the sublattice diagonal disorder.
After all the $V_n$ is non-zero only if $n$ is odd, i.e.
$V_{2n-1}$ is random and $V_{2n}=0$. Sometime the model is called the  
"periodic random alloy" or the "dilute Anderson model" \cite{milke97,biswas00}.
 Although in our case the energy at the deterministic even sites are zero, 
 the case where the deterministic sites
are non-zero but constant is trivially obtained by shifting the energy.  
The extended state at $E=0$ appears in the 
one-dimensional periodic random binary alloys
 due to a resonance \cite{wu91,biswas00}.
 It is demonstrated that the occurrence of the extended states in the vicinity 
of the site energy of the deterministic part, i.e. 
$E=0$ in our case. 
The delocalization can be understood in terms of the long-range 
correlation due to the perfect periodicity in the deterministic sites.

 In addition, we introduce another long-range correlation 
due to the base-pair sequence on the odd sites $V_{2n-1}$.
  Figures 7 and 8 show the energy dependence of the Lyapunov exponents 
($\gamma_1$ and $\gamma_2$)  
in some correlated cases ($B=1.0, 1.5, 1.8$). 
It found that the correlation enhances 
the localization length $\xi $($\equiv \gamma_2^{-1}$)  
around $|E|< 1$, although the largest Lyapunov exponent $\gamma_1$ is almost remained.

In Fig.9(a) the localization length $\xi_{d}=1/\gamma_d $ defined by the 
least nonnegative Lyapunov exponent is shown. 
  The electronic states whose energy is close to the resonance turn out to be 
extended in the sense that the localization length is the same order or 
larger than the system size. 
Accordingly the correlation effect is not so strong on the global feature
in the energy dependence of the density of states, particularlly 
the correlation enhances the localization length 
 in the energy regime $|E|<1$.

\begin{figure}[h]
\includegraphics[scale=0.7]{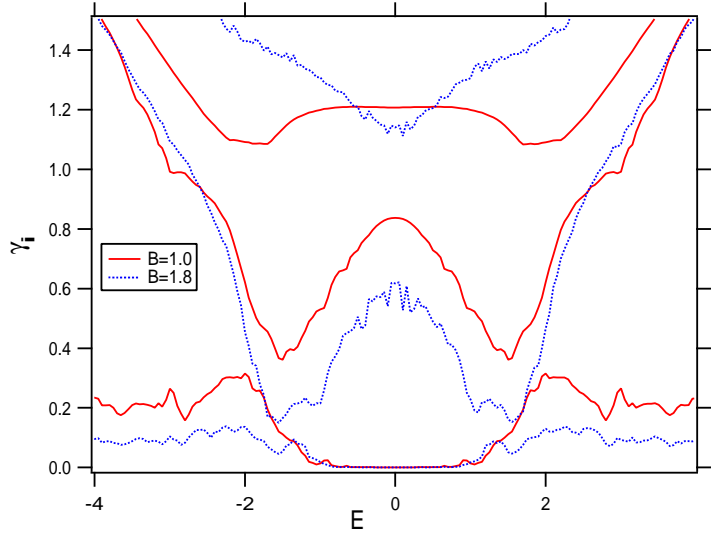}
  \caption{Localization length  $\xi(=1/\gamma_2)$
  as a function of energy in the correlated (a)two-chain model
and (b)three-chain model.
The parameters $W_{AT}=2.0, W_{AT-AT}=1.0, a=1.0, b=0.5$.
The on-site energy is set at $A_{nn}=B_{nn}=0$.
The number of the matrix product is $N=10^6$.}
\end{figure}

\subsection{Three-chain model}
 In this subsection, we confirm the numerical results for the three-chain
model when compared to that in the two-chain model.
 It must be noted that if we take into account of a $\pi-\pi$ interaction between 
the stacked base pairs, the model system becomes three chains model.
 (See Fig.2(b).)

As has been done
 in the two-chain case we can rewrite the Schr\H{o}dinger equation in the matrix
form as follows, 
 \begin{eqnarray}  
\left(
\begin{array}{c}
\phi^{A}_{n+1}+  \phi^{A}_{n-1} \\
\phi^{C}_{n+1}+  \phi^{C}_{n-1} \\
\phi^{B}_{n+1}+  \phi^{B}_{n-1} 
\end{array}
\right)
=
\left(
\begin{array}{ccc}
E-A_{nn} & -V_n & 0 \\
-\frac{V_n}{t_c} & \frac{E-C_{nn}}{t_c} & \frac{U_n}{t_c} \\
0 & U_n & E-B_{nn}
\end{array}
\right)
\left(
\begin{array}{c}
\phi^{A}_{n} \\
\phi^{C}_{n} \\
\phi^{B}_{n} 
\end{array}
\right) \equiv 
R
\left(
\begin{array}{c}
\phi^{A}_{n} \\
\phi^{C}_{n} \\
\phi^{B}_{n} 
\end{array}
\right) 
.
\end{eqnarray}  
\noindent
 We set $a=b=1$, $C_{nn+1}=t_{c}$ for simplicity.
It should be noted that the matrix $R$
can not be diagonalized by any unitary transform because
the matrix is not a normal matrix, $R^tR \neq RR^t$, 
when $t_c \neq 1$ (Toeplitz theorem).
 Accordingly the simple expectation in the previous subsection 
is not suitable for the three-chain model.

Figures 10 and 11 show the energy dependence of the Lyapunov exponents 
in the three chain cases. 
 We can observe that all the Lyapunov exponents $\gamma_i(i\leq d) $ are changed 
by the correlation. The least nonnegative Lyapunov exponent $\gamma_3$ is diminished by
the correlation. 
The localization length are shown in Fig.9(b).
 We can see that the localization lenght diverges at the band center $E=0$.

  Generally speaking, in the quasi-one-dimensional chain 
with the hopping disorder 
the singularity of DOS, the localization length and the 
conuctance at the band center depend on the parity, bipartiteness 
and the boundary condition \cite{brouwer02,biswas00}. 
As the discussion about the details is out of scope 
of this paper, we give the simple comment.    
Note that the parity effects appear in the odd number chain 
with the hopping randomness.
In the odd number chain with a hopping randomness, only one mode at $E=0$ 
is remained as the extended state, i.e. $\gamma_d=0 $, and the
other exponents are positive, $\gamma_{d-1} > \dots > \gamma_1 >0$.
 The behavior is seen in Fig.10 and 11.
 Then,  
 the non-localized states with $\gamma =0$ determine the 
conductance.
  Although we ignored the bipartite structure in the three-chain model for simplicity, 
if we introduct the bipartiteness in the intrachain hopping integral $V_n(=U_n)$ 
the other delocalized state due to the chiral symmetry appears at $E=0$
as seen in the ladder model.   

Here, we stress again that in the more realistic model of DNA 
 the parity effects and the chiral symmetry are broken down by 
the  on-site energy fluctuation.
Indeed, Fig.12(a) shows the energy dependence of the Lyapunov exponents
in the system which the on-site energy $A_{nn}(=B_{nn})$ in the side chains
is also randomly genarated as well as the hopping disorder in Fig.11.
 In this case we can confirm that
 the least nonnegative Lyapunov exponent, $\gamma_3$, is positive.
It is found that the extended state at $E=0$ disappears and the localization length
becomes finite. (See Fig.12(b).)
  As a result, the localization properties can be changed by the introduction
of the correlation and the fluctuation of the sequence even in scope of our models.

\begin{figure}[h]
\includegraphics[scale=0.7]{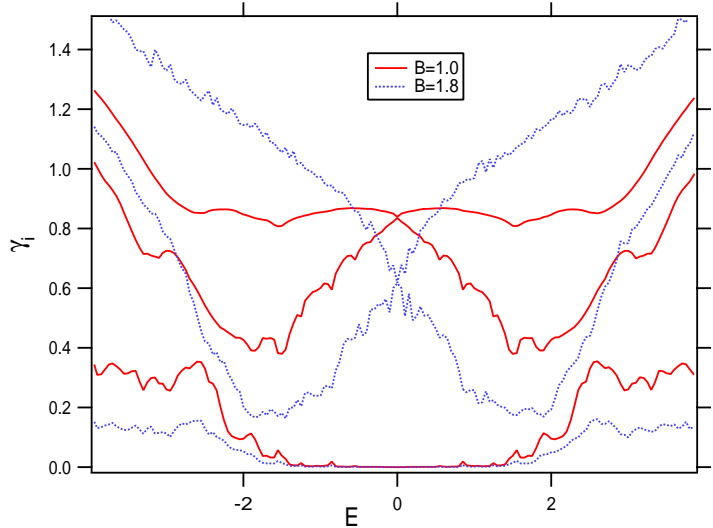}
  \caption{Lyapunov exponents $\gamma_i (i=1,2,3)$
  as a function of energy in the correlated three-chain model.
The parameters $W_{AT}=2.0, W_{AT-AT}=1.0, a=1.0, b=1.0$.
The on-site energy is set at $A_{nn}=B_{nn}=C_{nn}=0$.}
\end{figure}

\begin{figure}[h]
\includegraphics[scale=0.7]{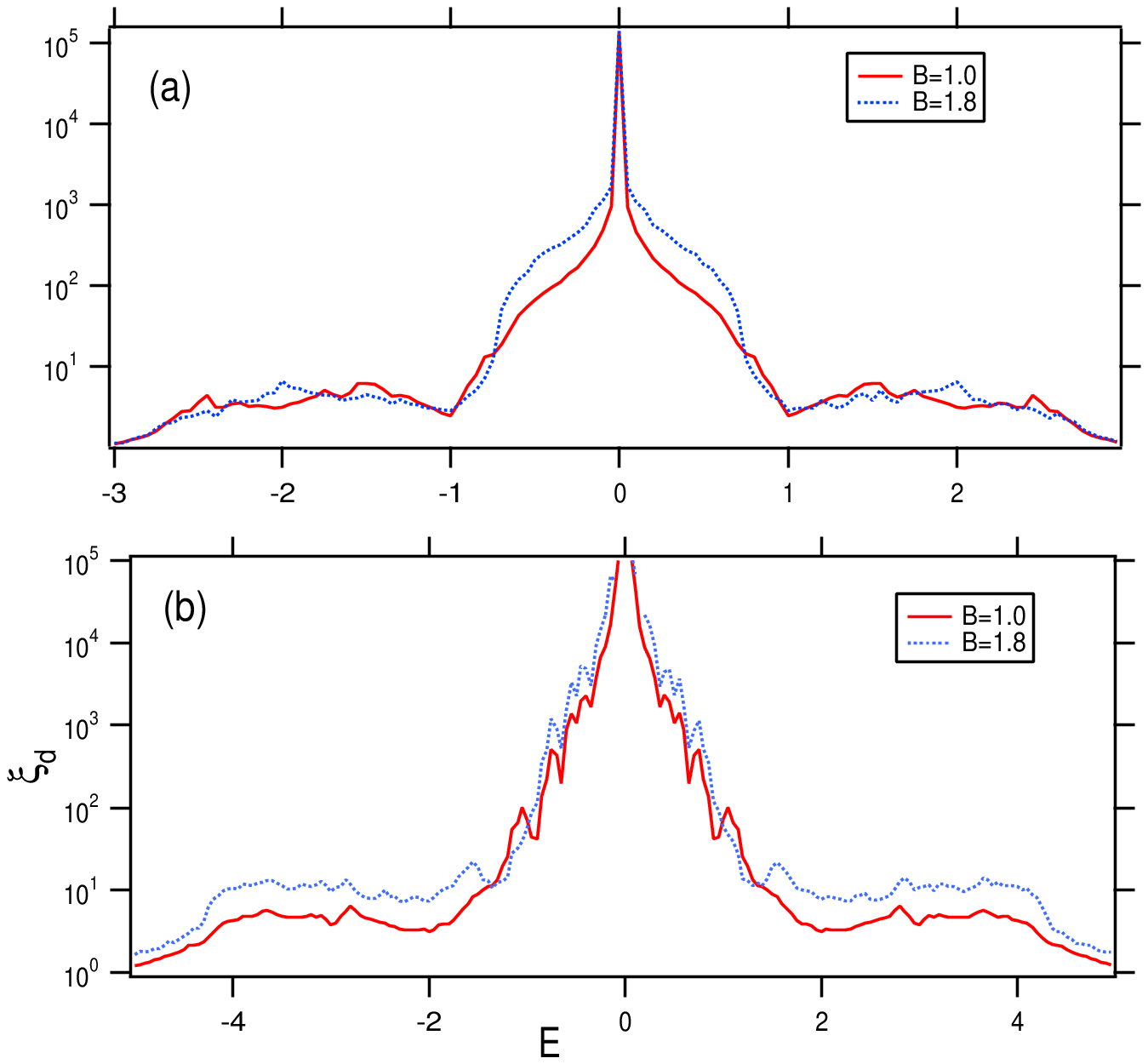}
  \caption{Lyapunov exponents $\gamma_i (i=1,2,3)$
  as a function of energy in the correlated three-chain model.
The parameters $W_{AT}=2.0, W_{AT-AT}=1.0, a=1.0, b=0.5$.
The on-site energy is set at $A_{nn}=B_{nn}=C_{nn}=0$.
}
\end{figure}

\begin{figure}[h]
\includegraphics[scale=0.7]{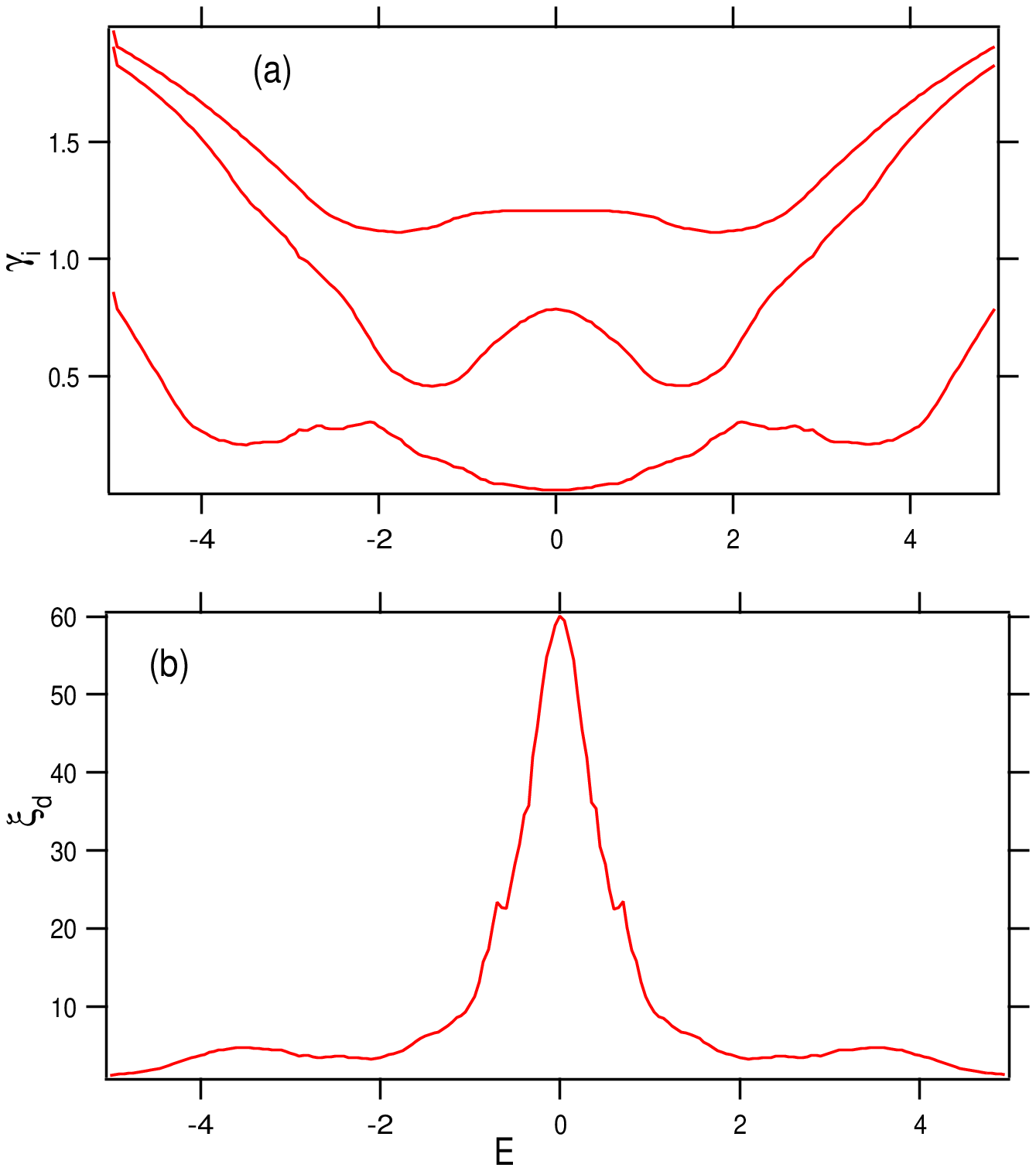}
  \caption{(a) Lyapunov exponents $\gamma_i (i=1,2,3)$
  as a function of energy in the disordered three-chain model ($B=1.0$).
(b) The localization length defined by $\xi_d\equiv 1/\gamma_3$.
The parameters $W_{AT}=2.0, W_{AT-AT}=1.0, a=1.0, b=0.5$ and $C_{nn}=0$.
The on-site energy of the side chains, $A_{nn}(=B_{nn})$, 
is randomly generated.
}
\end{figure}

\section{Summary and Discussion}
We numerically investigated the correlation effect on 
the localization property of the one-electronic states 
in the disordered two-chain (ladder) and three-chain models  
with the long-range structural correlation
as the simple models for electronic property in the double strand of DNA.
The results are summarized as follows.

(1) We gave 
  the tight-binding models for 
the few-chain systems with the off-diagonal randomness as the interchain 
and/or the intrachain hopping integrals.

(2) A simple interpretation for the results obtained in 
the two-chain model was given by means of the unitary transform.

(3) The correlation enhances the localization length ($\gamma_2^{-1}$)  
around $|E|<1$, although the $\gamma_1$ is almost remained, in the two-chain model.

(4) In the three-chain model, all the Lyapunov exponents are changed 
by the correlation effect, particularlly, the least nonnegative exponent  
$\gamma_3$ is diminished by the long-range correlation. 

(5) The divergence of the localization length around the $E=0$ 
disappears by the introduction of the fluctuation of the on-site energy.

The details of the energy dependence, the correlation dependence and the potential 
strength dependence on the DOS and the Lyapunov exponents, 
will be given elsewhere \cite{yamada04a}.

  Although we concentrated on the localization property of the correlated 
system for the double strand of DNA by using simple values
for the parameters, 
the more realistic Hamiltonian matrix elements are obtained from 
the oligomer calculation using the extended H\H{u}ckel theory
with the Clementi's parametization \cite{ladik88,iguchi03a,roche03,roche03a}.

In the experiment of the conductance property of the DNA both the temperature
dependence and the temperature effect are important.
Finite temperature can also reduce the effective system size and leads to 
the changes in the transport property.
Moreover, the effects of the stacking energy and of temperature 
can be considered by introducing the fluctuation in the hopping
energy such 
as the Su-Schriefer-Heegar model for polyacetilene \cite{su80}. 
 It is interesting to investigate the quantum diffusion
in order to reveal the conducting properties
of electrons, polarons and solitons in the fluctuating
ladder models \cite{yamada04a}.
  In the present model, although we used the binary correlated sequence,
the four letter virsion (A,T,G,C) is also interesting \cite{iguchi94}.
An extension for the four letter virsion with a long-range
correlated hopping disorder has been constructed \cite{yamada04a}.

\begin{acknowledgments}
The author would like to thank Dr. Kazumoto Iguchi for 
stimulating and useful discussions and providing his 
papers and preprints.
 The author also would like to thank Kazuko Iguchi for kind
hospitality during the stay in Anan where he got a good chance 
to start this study.
\end{acknowledgments}


\end{document}